\documentclass[prd,twocolumn,showpacs,aps,superscriptaddress]{revtex4}
\usepackage{graphicx}

\begin{document}

\title{Destruction of small-scale dark matter clumps \\
in the hierarchical structures and galaxies}

\author{Veniamin Berezinsky}
 \email{berezinsky@lngs.infn.it}
 \affiliation{INFN, Laboratori Nazionali del Gran Sasso, I-67010
  Assergi (AQ), Italy}
 \affiliation{Institute for Nuclear Research of the Russian Academy of
 Sciences, Moscow, Russia}
\author{Vyacheslav Dokuchaev}
 \email{dokuchaev@inr.npd.ac.ru}
\author{Yury Eroshenko}
 \email{erosh@inr.npd.ac.ru}
 \affiliation{Institute for Nuclear Research of the Russian Academy of
 Sciences, Moscow, Russia}

\date{\today}

\begin{abstract}
A mass function of small-scale dark matter clumps is calculated in the
standard cosmological scenario with an inflationary-produced primordial
fluctuation spectrum and with a hierarchical clustering. We take into
account the tidal destruction of clumps at early stages of structure
formation starting from a time of clump detachment from the Universe
expansion. Only a small fraction of these clumps, $\sim0.1-0.5$~\%, in
each logarithmic mass interval $\Delta\log M\sim1$ survives the stage
of hierarchical clustering. The surviving clumps can be disrupted
further in the galaxies by tidal interactions with stars. We performed
the detailed calculations of the tidal destruction of clumps by stars
in the Galactic bulge and halo and by the Galactic disk itself. It is
shown that the Galactic disc provides the dominant contribution to the
tidal destruction of small-scale clumps outside the bulge. The results
obtained are crucial for calculations of the dark matter annihilation
signal in the Galaxy.
\end{abstract}

\maketitle

\section{Introduction}

One of the promising indirect manifestation of the Dark Matter
(DM) particles is their possible annihilation in the Galactic halo
\cite{SiBl}. A local annihilation rate is proportional to the
square of number density of DM particles. Therefore, annihilation
proceeds more efficiently in the dense DM substructures of the
Galactic halo. Both analytical calculations
\cite{silk93,ufn2,bde03,BBDG05} and numerical simulations
\cite{moore99,ghigna98,klypin99} with the inflationary-produced
adiabatic density fluctuations predict the existence of DM clumps
in the Galactic halo. The enhancement of the annihilation signal
due to the presence of substructures in the Galactic halo depends
on the fraction of the most dense small-scale clumps
\cite{bde03,moore05}. The most essential characteristics of clumps
for calculations of DM annihilation in the Galactic halo are the
minimum mass and distribution function of clumps. At the same time
the tidal destruction of clumps \cite{bde03} strongly influences
the number density of clumps in the Galaxy.

The small-scale clumps  \cite{silk93,ufn2,bde03} are formed only if the
corresponding density fluctuations are large enough. The inflation
models predict the power-law primordial fluctuation spectrum with an
power index $n_p \approx 1.0$. The small-scale clumps are formed
earlier than the larger ones and captured by the larger clumps in the
process of a hierarchical clustering in the expanding Universe.
Eventually all clumps consist in part of the smaller ones and of the
free DM particles. An effective index of the density perturbation power
spectrum $n\to -3$ at small-scales (when mass inside the perturbation
$M\to 0$). This means that a gravitational clustering of small-scale
structures proceeds very fast. As a result the formation of new clumps
and their capturing by the larger ones are nearly simultaneous
processes.

A convenient formalism, which describes statistically this hierarchical
clustering, is the Press-Schechter theory \cite{ps74} and its
extensions, in particular the `excursion set' formalism developed by
Bond et al. \cite{bond91} (for a clear introduction see \cite{cole}).
However, this theory does not include an important process of the tidal
destruction of small clumps inside the bigger ones. This process has
been taken into account in our previous work \cite{bde03}, where it was
demonstrated that only a small fraction of the small-scale clumps
survives the tidal destruction in the hierarchical clustering.
Nevertheless, even this small fraction of survived small-scale clumps
is enough to dominate the DM annihilation rate for the most reasonable
spectra of primordial fluctuations.

A mass distribution of small-scale clumps survived in the hierarchical
structuring was derived in \cite{bde03}:
\begin{equation}
 \xi_{\rm int}\, \frac{dM}{M} \simeq 0.01 (n+3)\, \frac{dM}{M} ,
 \label{xi1int}
\end{equation}
where $M$ is a clump mass, $n$ is a power-index of density
perturbations at a mass-scale $M$. The distribution function $\xi_{\rm
int}$ is a mass fraction of DM in the form of clumps in the logarithmic
mass interval $d\log M$.

The minimal mass of DM clumps $M_{\rm min}$ is determined by the
leakage of DM particles from the growing density fluctuations (the
diffuse leakage and free streaming) and depends on the properties of DM
particles \cite{fsfirst,ufn2,bino,Schw03,GHS04,GHS0508,GHS0503}. The
existing estimates of $M_{\rm min}$ for neutralino DM are substantially
different, from $M_{\rm min}\sim 10^{-12}M_{\odot}$ in \cite{gzv12} to
$M_{\rm min}\sim(10^{-7}-10^{-6})M_{\odot}$ in
\cite{bino,GHS04,GHS0503}. In \cite{bde03} we performed detailed
calculations of a DM particle diffusion and free streaming in the
kinetic equation approach. For the case of neutralino considered as a
pure bino we obtained for the minimal mass of DM clumps
\begin{eqnarray}
M_{\rm min}&=&1.5\times10^{-8}\left(\frac{m_{\chi}}{100\mbox{
GeV}} \right)^{-15/8} \left(\frac{\tilde M}{1 \mbox{
TeV}}\right)^{-3/2}
\nonumber\\
&&\times\left(\frac{g_*}{10}\right)^{-15/16}
\left(\frac{\Lambda^*}{83}\right)^3 M_{\odot}, \label{mminnum}
\end{eqnarray}
where $\tilde{M}^2=\tilde{m}^2-m_{\chi}^2$, with $m_{\chi}$ being a
neutralino mass, $\tilde{m}$ is a sfermion mass, and $\Lambda^*$ has
only a logarithmic dependence on $\tilde M$ and $m_{\chi}$. In the
considered range of parameters $\Lambda^*\simeq83$. Our value of
$M_{\rm min}$ agrees reasonably well with \cite{bino,GHS04,GHS0503} and
strongly disagrees with \cite{gzv12} for the reasons explained in
\cite{bde03}.

Due to uncertainties in the SUSY parameters, a numerical value of
$M_{\rm min}$ is not exactly predicted. With our choice of SUSY
parameters \cite{bde03}, $M_{\rm min}\sim10^{-8}M_{\odot}$, is of the
Moon-scale mass. With other choice of SUSY parameters \cite{GHS0503},
$M_{\rm min}\sim10^{-6}M_{\odot}$, is of the Earth-scale mass. In the
numerical simulations \cite{DieMooSta05}, $M_{\rm
min}\sim10^{-6}M_{\odot}$, was actually assumed by putting the
corresponding cut-off in the initial density perturbations.

The very interesting numerical simulations of the formation of
small-scale DM clumps with a mass larger than the Earth mass have been
performed recently in \cite{DieMooSta05}. There is direct
correspondence of this simulations with the earlier theoretical
calculations:

(i) The density profile of large-scale clumps is influenced by the
hierarchical clustering of the smaller ones. The new important
result of numerical simulations \cite{DieMooSta05} is a resolution
of the density profile of the isolated  minimal mass clumps,
$M_{\rm min}\sim10^{-6}M_{\odot}$. The clumps of minimal mass are
formed directly from the isolated fluctuations and their density
profile is not influenced by the hierarchical clustering. The
internal density profile of small-scale clumps in these
simulations is proved out the same as in the theoretical
calculations \cite{ufn1} performed for the isolated density
fluctuations. The agreement between the theory and numerical
simulations for the predicted internal density profile of clumps,
$\propto r^{-\beta}$, is fairly good within the involved
uncertainties: $\beta=1.7-1.8$ in \cite{ufn1}, and $\beta=1.5-2.0$
in \cite{DieMooSta05} respectively.

(ii) The numerical simulations \cite{DieMooSta05} agree rather
well with the shape of theoretically derived mass function of
small-scale clumps \cite{bde03} but with the different
normalizations.

A tidal destruction in the Galaxy of the Earth-size clumps from
simulations \cite{DieMooSta05} has been recently considered in
\cite{ZTSH,ZTSH0508,MDSQ}. The results are rather controversial.
Authors of \cite{ZTSH,ZTSH0508} conclude that all the Earth-mass clumps
are destructed in tidal interactions with stars in the Galaxy, while in
\cite{MDSQ} this result was not confirmed under a different assumption
on the star number density.

In this paper we present the alternative and independent calculations
for all processes of the tidal destruction of small-scale clumps: (i)
in the hierarchical clustering, (ii) by stars from the stellar bulge,
(iii) by stars from the halo and (iv) by the Galactic disc. The last of
these processes is turned out the most effective. We also describe a
new method for calculations of clump destruction in the hierarchical
clustering, which is a more general (valid for the arbitrary spectra of
primeval fluctuations) and formally more transparent than the earlier
one in \cite{bde03}.

Our calculations of the tidal destruction of clumps by stars in
the Galaxy are quite different from \cite{ZTSH,ZTSH0508,MDSQ} by
both, methods and results. While in the references above, only the
tidal destructions in collisions of clumps with the individual
stars were studied, we found that dominant effect is provided by
the destruction of clumps in the collective gravitational field of
the Galactic disc. As a result we predict that only 17\% of the
Earth-mass clumps survived the tidal destruction at the position
of the Sun. This result is crucial for the rate of DM annihilation
in the Galaxy.

The paper is organized as it follows: In the Sec.~\ref{hdestr} we
describe a new method for calculation of the small-scale clump
destruction in the hierarchial clustering. We calculate a mass function
of survived clumps and compare it with a corresponding one from
numerical simulations. In the Sec.~\ref{disksec} the tidal destruction
of clumps by the Galactic disk is considered. In the Sec.~\ref{stars}
the life-time of clumps in the central stellar bulge and stellar halo
spheroid is calculated. In the Sec.~\ref{discussion} we discuss the
obtained results.

We perform our calculations for the standard cosmological model
with a matter density parameter $\Omega_{\mathrm{m}}=0.3$, a
cosmological constant term $\Omega_{\Lambda} =
1-\Omega_{\mathrm{m}}\simeq0.7$ and the Hubble constant $h=0.7$.

\section{Destruction of clumps in hierarchial clustering}
\label{hdestr}

The process of hierarchical clustering and tidal destruction of DM
clumps can be outlined in the following way. The DM clumps of minimal
mass are formed first in the expanding Universe. The clumps of larger
mass, which host the smaller ones are formed later, and so on. Some
part of clumps are destroyed tidally in the gravitational field of
their host clumps.

In this Section we study the destruction of DM clumps in the process of
hierarchical structuring long before the final galaxy formation. At
small-mass scales the hierarchial clustering is a fast and rather
complicated nonlinear process. We use a simplified model which
nevertheless takes into account the most important features of
hierarchial clustering.

To describe the formation of clumps we will use the model of
spherical collapse \cite{cole} in flat cosmology without the
$\Lambda$-term. This assumption is well justified at early times
of clumps formation when the $\Lambda$-term is negligible in
comparison with the matter density. In this model a formation time
of clump with an internal density $\rho$ is $t=(\kappa \rho_{\rm
eq}/\rho)^{1/2}t_{\rm eq}$, where $\kappa=18\pi^2$ and $\rho_{\rm
eq}=\rho_0(1+z_{\rm eq})^3$ is a cosmological density at the time
of matter-radiation equality $t_{\rm eq}$, $1+z_{\text{eq}}=
2.35\times10^4\Omega_mh^2$ and $\rho_0=
1.9\times10^{-29}\Omega_mh^2\mbox{ g cm}^{-3}$. The index `eq'
here and throughout below refers to quantities at the time of
matter-radiation equality $t_{\rm eq}$.

The DM clumps of mass $M$ can be formed from density fluctuations
of different peak-height $\nu=\delta_{\rm eq}/\sigma_{\rm eq}(M)$,
where $\sigma_{\rm eq}(M)$ is the fluctuation dispersion on a
mass-scale $M$ at the time $t_{\rm eq}$. A mean internal density
of clump $\rho$ is fixed at the time of clump formation and
according to \cite{cole} equals $\rho= \kappa\rho_{\rm
eq}[\nu\sigma_{\rm eq}(M)/\delta_c]^3$, where
$\delta_c=3(12\pi)^{2/3}/20\simeq1.686$.

The tidal destruction of clumps is most effective at the early epochs
of the Galactic halo formation, when the host density profiles are not
finally established. The tidal interaction of clumps is a complicated
process and depends on many factors: a clump formation history, host
density profile, an existence of different substructures inside the
host, orbital parameters of individual clumps in the hosts, etc. Only
in numerical simulations all these factors can be taken into account
properly. In this paper we use a simplified approach by calculating an
energy gain per each tidal interaction and a number of tidal
interactions per dynamical time in the hosts.

An internal energy of self-gravitating object increases in tidal
interactions. This energy increase was calculated e.~g. in
\cite{gnedin1} for the case of a star globular cluster in a spherical
galaxy. By using the model of tidal heating from \cite{gnedin1}, we
determine now a survival time (or a time of tidal destruction) $T$ of
some chosen small-scale clump due to the tidal heating inside of a
larger mass host clump. The motion of a clump would be rather
complicated in the case of a fast hierarchical clustering of hosts.
During a dynamical time in the host $t_{\rm
dyn}\simeq0.5(G\rho_h)^{-1/2}$, where $\rho_h$ is a mean internal
density of the host, the chosen small-scale clump can belong to several
successively destructed hosts. We will consider a typical clump orbit
inside the host and assume for simplicity in this Section the
isothermal internal density profile of the clump.

A clump trajectory in the host experiences successive turns accompanied
by the ``tidal shocks'' \cite{spit,gnedin1}. For the considered
small-scale clump with a mass $M$ and radius $R$, the corresponding
internal energy increase after a single tidal shock is
\begin{equation}
 \Delta E\simeq\frac{4\pi}{3}\,\gamma_1G\rho_hMR^2,
 \label{dele}
\end{equation}
where a numerical factor $\gamma_1\sim1$. Let us denote the number
of tidal shocks per dynamical time $t_{\rm dyn}$ by $\gamma_2$. A
corresponding rate of clump internal energy growth is $\dot
E=\gamma_2\Delta E/t_{\rm dyn}$. A clump is destroyed in the host
if its internal energy increase due to tidal shocks exceeds a
total energy $|E|\simeq GM^2/2R$. As a result, for a typical time
$T=T(\rho,\rho_h)$ of the tidal destruction of a small-scale clump
with density $\rho$ inside a more massive host with a density
$\rho_h$ we obtain:
\begin{equation}
T^{-1}(\rho,\rho_h)=\dot E/|E|\simeq4\gamma_1\gamma_2
G^{1/2}\rho_h^{3/2}\rho^{-1}.
\end{equation}
It turns out that a resulting mass function of small-scale clumps (see
in this Section below) depends rather weakly on the value of
$\gamma_1\gamma_2$.

During the lifetime of an individual small-scale clump, it can
sequentially inhabit in many host clumps of larger mass. After the
tidal disruption of the first lightest host, a small-scale clump
becomes a constituent part of a heavier one, etc. The process of
hierarchical transition of a small-scale clump from one host to another
occurs almost continuously in time up to the final host formation,
where the tidal interaction becomes inefficient.

A corresponding mass fraction of small-scale clumps with mass $M$
escaping the tidal destruction in hierarchical clustering (or
probability of clump survival) is given by the exponent function
$e^{-J}$ with
\begin{equation}
 J\simeq\sum\limits_{h} \frac{\Delta t_h}{T(\rho,\rho_h)}.
 \label{jsum}
\end{equation}
Here $\Delta t_h$ is a difference of formation times $t_h$ of two
successive hosts, and summation is over all clumps of intermediate
mass-scales, which successively host the chosen small-scale clump of a
mass $M$. By changing the summation to integration in (\ref{jsum}) we
obtain
\begin{equation}
 J(\rho,\rho_f)=\int\limits_{t_1}^{t_f}\!\frac{dt_h}{T(\rho,\rho_h)}
 \simeq\gamma\frac{\rho_1-\rho_f}{\rho}
 \simeq\gamma\,\frac{\rho_1}{\rho}\,,
 \label{sumint1}
\end{equation}
where
\begin{equation}
 \gamma=2\gamma_1\gamma_2\kappa^{1/2}G^{1/2}\rho_{\rm eq}^{1/2}t_{\rm eq}
 \simeq14(\gamma_1\gamma_2/3),
 \label{bigj14}
\end{equation}
and $t_1$, $t_f$, $\rho_1$ and $\rho_f$ are respectively the formation
times and internal densities of the first and final hosts. It can be
seen from (\ref{sumint1}) that the first host provides a major
contribution to the tidal destruction of small-scale clumps, especially
if the first host density $\rho_1$ is close to $\rho$, and consequently
$e^{-J}\ll 1$.

A mass function of small-scale clumps (i.~e. a differential mass
fraction of DM in the form of clumps survived in hierarchical
clustering) can be expressed as
\begin{eqnarray}
 \xi\frac{dM}{M}\,d\nu &=& dM\,d\nu\,(2/\pi)\,e^{-\nu^2/2}\!
 \int\limits_0^{\nu}d\nu_1\,e^{-\nu_1^2/2} \nonumber \\
&& \times  \!\!\int\limits_{t_1(\nu_1)}^{t_0}\!d\tilde t \,\left|
\frac{\partial^2 F(M,\tilde t)}{\partial M~\partial \tilde t} \right|
 e^{-J[\rho(\nu),\rho(\,\tilde{t}\,)]}.
\label{phiin}
\end{eqnarray}
In this expression $t_0$ is the present age of the Universe and
$F(M,t)$ is the mass fraction of unconfined clumps (i.~e. clumps
not belonging to the more massive hosts) with a mass smaller than
$M$ at time $t$. According to \cite{cole}, the mass fraction of
unconfined clumps is $F(M,t)={\rm erf}(
\delta_c/[\sqrt{2}\sigma_{\rm eq}(M)D(t)])$, where ${\rm erf}(x)$
is the error-function and $D(t)$ is the growth factor normalized
by $D(t_{\rm eq})=1$. An upper integration limit $t_0$ in
(\ref{phiin}) is not crucial and may be extrapolated to infinity
because a main contribution to the tidal destruction of clumps is
provided by the early formed hosts at first steps of hierarchical
clustering.

Two processes respond for a time evolution of the fraction $\partial
F(M,t)/\partial M\,dM$ of unconfined clumps in the mass interval $dM$:
(i) the formation of new clumps and (ii) the capture of smaller clumps
into the larger ones. Both these processes are equally effective at the
time when $\partial^2 F/(\partial M\partial t)=0$. To take into account
the confined clumps (i.~e. clumps in the hosts) we need only the 2nd
process (ii) for the fraction $\partial F(M,t)/\partial M$.
Nevertheless, in (\ref{phiin}) it is used the fraction $\partial
F(M,t)/\partial M$ influenced by the both processes. This is not
accurate at a typical formation time of clump with a mass $M$, when
clump density is comparable with the density of hosts. Fortunately, for
this time the exponent in (\ref{phiin}) is very small, $e^{-J}\ll 1$,
as it can be seen from (\ref{sumint1}) and (\ref{bigj14}).
Respectively, an uncertain contribution from the process (i) to the
integral (\ref{phiin}) is also very small. Meanwhile, only the process
(ii) dominates in the integration region where the exponent $e^{-J}$ is
not small. For this reason (\ref{phiin}) provides a suitable
approximation for the mass fraction of clumps survived in the
hierarchical clustering.

\begin{figure}[t]
\includegraphics[width=0.48\textwidth]{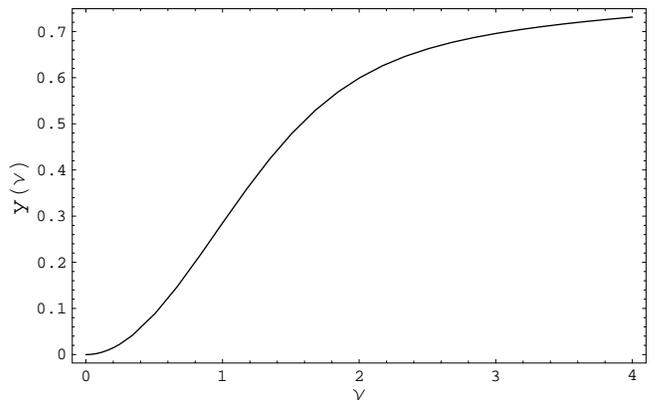}
\caption{Numerically calculated function $y(\nu)$ from
(\ref{psiitog}).}
 \label{funfirst}
\end{figure}

Finally we transform the distribution function (\ref{phiin}) to the
following form:
\begin{equation}
 \xi\,\frac{dM}{M}\,d\nu\simeq
 \frac{1}{\sqrt{2\pi}}\,e^{-\nu^2/2}y(\nu)\,d\nu\,
 \frac{d\log\sigma_{\rm eq}(M)}{dM}\,dM.
 \label{psiitog}
\end{equation}
Here the numerically calculated function $y(\nu)$ depends rather weakly
on the parameter $\gamma$ from (\ref{bigj14}) and is shown in the
Fig.~\ref{funfirst}. By deriving (\ref{psiitog}), we take into account
that $\sigma(M)$ is a slowly varying function of clump mass $M$. For
the same reason, providing an integration in (\ref{phiin}) we use the
dependance of $t_1(\rho_1)$ only on the variable $\nu$ by neglecting
the dependance on $M$. Physically the rising of $y(\nu)$ with $\nu$
corresponds to a more effective survival of high-density clumps (i.~e.
with large values of $\nu$) with respect to the low-density ones (with
small values of $\nu$). Integrating (\ref{psiitog}) over $\nu$, we
obtain
\begin{equation}
 \xi_{\rm int}\frac{dM}{M}\simeq0.017(n+3)\,\frac{dM}{M}.
 \label{xitot}
\end{equation}
This mass function is in a reasonable agreement with the similar one
from (\ref{xi1int}) in the our earlier work \cite{bde03}. An effective
power-law index $n$ in (\ref{xitot}) is given by
\begin{equation}
  n=-3\left[1+
  2\frac{\partial\log\sigma_{\rm eq}(M)}{\partial\log M}\right]
  \label{neff}
\end{equation}
and depends very weakly on $M$. At the small mass-scales one has
approximately $n\simeq n_p-4$. Equation (\ref{xitot}) implies that for
the suitable values of $n$ only a small fraction of clumps, about
$0.1-0.5$~\%, survives the stage of hierarchical tidal destruction in
the each logarithmic mass interval $\Delta\log M\sim1$. A simple
$M^{-1}$ shape of the mass function (\ref{xitot}) is in a very good
agreement with the corresponding one obtained recently in the numerical
simulations \cite{DieMooSta05}.

It must be stressed that a physical meaning of the survived clump
distribution function $\xi\,d\nu\,dM/M$ is different from the similar
one for the unconfined clumps, given by the Press-Schechter mass
function $\partial F/\partial M$. For comparison, the Press-Schechter
mass function of unconfined clumps \cite{cole} is
\begin{equation}
 \xi_{\rm PS}(t)\,\frac{dM}{M}=\frac{2\delta_c}{\sqrt{2\pi}
 \sigma_{\rm eq}^2D(t)}\frac{d\sigma_{\rm eq}}{dM}
 \exp\!\left[-\frac{\delta_c^2}{2\sigma_{\rm eq}^2D^2(t)}\right]dM,
 \label{xips}
\end{equation}
where $\sigma_{\rm eq}=\sigma_{\rm eq}(M)$ The mass function of clumps
survived in a hierarchical clustering (\ref{xitot}) is several times
less than the Press-Schechter mass function (\ref{xips}) at a mean time
of clump formation with $\sigma_{\rm eq}(M)D(t)\simeq\delta_c$.

In further calculations we will use an interpolation fitting of the
fluctuation dispersion $\sigma_{\rm eq}(M)$ from \cite{bugaev} (see
also \cite{ssw}):
\begin{equation}
 \sigma_{\rm eq}(M)\simeq
 \frac{2\times10^{-4}}{\sqrt{f_s(\Omega_{\Lambda})}}
 \left(\frac{k}{k_{h0}}\right)^{\!(n_p-1)/2}
 \left[\log\left(\frac{k}{k_{\rm eq}}\right)\right]^{3/2}\!\!,
 \label{A5}
\end{equation}
where the wave vector $k\propto M^{-1/3}$, respectively $k_{\rm eq}$
and $k_{h0}$ correspond to a mass inside the cosmological horizon at
the moments $t_{\rm eq}$ and $t_{0}$, $n_p$ is a primordial
perturbation index, and $f_s(\Omega_{\Lambda})\!=\!1.04-0.82
\Omega_{\Lambda} +2\Omega_{\Lambda}^2$. It must be noted that
interpolation (\ref{A5}) is valid only for the small-scale clumps, with
$M\leq10^3M_{\odot}$. The analysis of the WMAP data of the CMB
anisotropy \cite{wmap} reveals a  power-law spectrum of initial
perturbations with $n_p=0.99 \pm 0.04$ in a good agreement with the
canonical inflation value $n_p=1.0$. However, when the data from 2dF
galaxy power-spectrum and Ly-$\alpha$ are included in the analysis, the
best-fit favors in a softer spectrum with $n_p=0.96 \pm 0.02$.
Nevertheless, the recent observations does not exclude even the values
$n_p=1.1$.

\begin{figure}[t]
\includegraphics[width=0.48\textwidth]{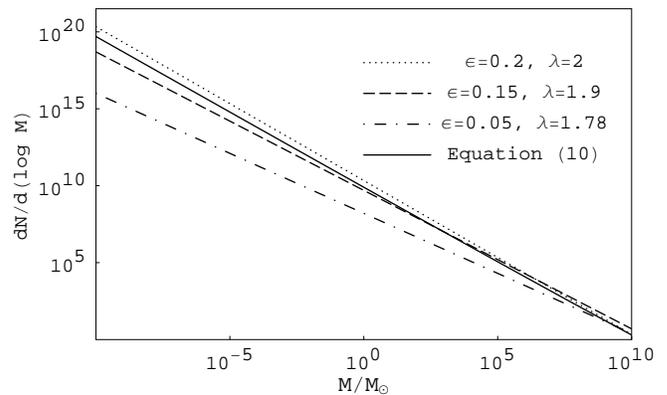}
\caption{A differential number of small clumps in the Galaxy from
(\ref{xitot}) for $n_p=1.0$ is shown by the solid line. These
calculations are valid only for small-scale clumps with
$M<10^3M_{\odot}$ and extrapolated to the larger masses (the right part
of the solid line). Other curves are the corresponding number densities
of large-scale clumps with $M>10^6M_{\odot}$ from numerical simulations
(see details in the text) for different values of parameters
$\varepsilon$ and $\lambda$ from (\ref{bigpsi}). The left parts of
these curves are the extrapolations to small masses.}
 \label{figsh}
\end{figure}

Note that a differential number density of small-scale clumps in the
Galactic halo $n(M)\,dM \propto dM/M^2$ from (\ref{xitot}) coincides
not only with a similar one from the recent numerical simulations of
small-scale clumps \cite{DieMooSta05} but is also very close to that
obtained in the numerical simulations for large-scale clumps with mass
$M\geq10^6M_{\odot}$. See Fig.~\ref{figsh} for a comparison.

Strictly speaking, our calculations are not valid for large-scale
clumps because of their continuing tidal destruction in the halo up to
the present epoch $t_0$ and the accretion of the additional large-scale
clumps into the halo from the intergalactic space. Nevertheless, our
approach remains valid even for the large-scale clumps in the narrow
mass range, where the power-law perturbation spectrum can be used as a
rather good approximation.

In the Fig.~\ref{figsh} a differential number density of small-scale
clumps from (\ref{xitot}) is shown by the solid line. As it was noted
above, the region of validity for this curve is $M\leq10^3 M_{\odot}$.
For larger masses an extrapolation is shown (right part of the solid
line). The corresponding mass functions from numerical simulations can
be parameterized in the form $\xi(M)\,dM/M=AM^{1-\lambda}\,dM$. The
constant $A$ can be determined by fixing a power-law index $\lambda$
and a fraction $\varepsilon$ of the halo mass in the form of clumps
with mass from $M_{\rm min}\simeq10^6M_{\odot}$ to $M_{\rm
max}\simeq10^{10}M_{\odot}$:
\begin{equation}
 \varepsilon=
 \int\limits_{M_{\rm min}}^{M_{\rm max}}AM^{1-\lambda}\,dM.
\end{equation}
With this parametrization a mass function of large-scale DM clumps may
be expressed as
\begin{equation}
 \xi(M)\,\frac{dM}{M}=\varepsilon\,\frac{dM}{M}\left\{
  \begin{array}{lcr}
 \displaystyle{\frac{(2-\lambda)M^{2-\lambda}}{{M_{\rm max}^{2-\lambda}} -
 {M_{\rm min}^{2-\lambda}}}}\, ,  & {} & \lambda\neq2; \\  \\
 \displaystyle{\log^{-1}\!\left(\frac{M_{\rm max}}{M_{\rm min}}\right)}\, ,
 & {} & \lambda=2.
  \end{array}
 \right.
 \label{bigpsi}
\end{equation}
In the Fig.~{\ref{figsh}} a differential number density of
large-scale clumps $(M_{\rm H}/M)\,\xi(M)\,dM/M$ from
(\ref{bigpsi}) is shown for different values of $\lambda$ and
$\varepsilon$ taken from various numerical simulations:
$\varepsilon\simeq0.2$, $\lambda=2$ from \cite{moore99};
$\varepsilon\simeq0.15$, $\lambda=1.9$ from \cite{CalMoo}, and
$\varepsilon\simeq0.05$, $\lambda=1.78$ from \cite{Sto03}.
Observations of the Galactic halo lensing \cite{clobs} give a
smaller clump fraction value, $\varepsilon\simeq0.02$. One can see
in the Fig.~\ref{figsh} a reasonable agreement between the
extrapolation of our calculations and the corresponding numerical
simulations of the large-scale clumps.

\section{Destruction of clumps by disk}
\label{disksec}

Crossing the Galactic disc, a clump can be tidally destructed by the
collective gravitational field of stars in the disc. This phenomenon is
similar to the destruction of a globular cluster by the ``tidal
shocking'' in the Galactic disc \cite{OstSpiChe}.

The rate of energy gain per unit mass due to the tidal shocking was
calculated in \cite{OstSpiChe}:
\begin{equation}
  \label{disksh1}
 \frac{d\tilde E}{dt}=\frac{4g_m^2(\Delta z)^2}{T_cv_{z,c}^2} .
\end{equation}
Here $g_m$ is the maximum gravitational acceleration acquired by the
constituent DM particle of the clump moving in the gravitational field
of the disk, $\Delta z=$ is a vertical (perpendicular to the disk
plane) distance of a DM particle from the clump center, $T_c$ is an
orbital period of clump in the halo, $v_{z,c}$ is a vertical velocity
of disk crossing. In (\ref{disksh1}) the two crossings of disk by a
globular cluster during the orbital period $T_c$ is assumed, while only
one disk crossing is typical in the case of elongated orbits of DM
clumps in the halo (see below).

A surface mass of the Galactic disk can be approximated by a simple
exponential law \cite{MarSuch}
\begin{equation}
  \label{diskmass}
\sigma_s(r)=\frac{M_d}{2\pi r_0^2}\,e^{-r/r_0},
\end{equation}
with $M_d=8\times10^{10}M_{\odot}$ and $r_0=4.5$~kpc. The maximum
gravitational acceleration during the disk crossing is
\begin{equation}
 g_m(r)=2\pi G\sigma_s(r).
 \label{diskacc}
\end{equation}
Following to \cite{ufn1}, we use a power-law parametrization of the
internal density profile of DM clumps:
\begin{equation}
 \rho_{\rm int}(r)=
 \frac{3-\beta}{3}\,\rho\left(\frac{r}{R}\right)^{-\beta},
 \label{rho}
\end{equation}
where $\rho$ and R is respectively a mean internal density and radius
of clump, $\beta=1.7-1.8$ and we put $\rho_{\rm int}(r)=0$ at $r>R$.
The corresponding power-law profile of small-scale clumps with
$\beta\simeq1.5-2$ has been recently found in numerical simulations
\cite{DieMooSta05}. For this profile a total (kinetic and potential)
energy of a clump is given by
\begin{equation}
 |E|=\frac{3-\beta}{2(5-2\beta)}\frac{GM^2}{R}.
 \label{Etot}
\end{equation}
Integrating (\ref{disksh1}) over the clump volume with the profile
(\ref{rho}), one can obtain a total rate of energy gain by clump
$dE/dt$ and then a time of clump destruction by the disk:
\begin{equation}
 t_d=\frac{|E|}{\dot E}=
 \frac{2(5-\beta)}{3(5-2\beta)}\frac{GT_c\rho v_{z,c}^2}{g_m^2}.
 \label{tdisk}
\end{equation}
Note that the adiabatic correction for the disk shocking (see e.~g.
\cite{GneOst96}) is very small in the case of DM clumps and may be
neglected.

To estimate the tidal shocking effect produced by different parts of
the Galactic disk at radial distance  $r$, let us consider at first a
toy halo model by assuming the circular orbits of DM clumps. Then a
disk crossing velocity $v_{z,c}$ equals to a circular velocity:
\begin{equation}
 v_{z,c}= v_{\rm rot}(r)=
 \left[\frac{GM_{\rm H}(r)}{r}\right]^{1/2},
\end{equation}
where $M_{\rm H}(r)$ is a halo mass inside the sphere of radius $r$.
Using for a clump orbital period $T_c=2\pi r/v_{\rm rot}(r)$, one
finally finds
\begin{equation}
 t_d=\frac{(5-\beta)}{3\pi(5-2\beta)}
 \frac{r_0^4\rho M_{\rm H}^{1/2}(r)r^{1/2}}{G^{1/2}M_d^2}\,e^{2r/r_0}.
 \label{tdisktoy}
\end{equation}
Comparison of a clump destruction time $t_d$ from (\ref{tdisktoy}) with
the Universe age $t_0$ shows that all clumps with the internal density
$\rho< 2\times10^{-22}$~g~cm$^{-3}$ are effectively destructed within
the radius $r<15$~kpc from the Galactic center. In particular, the
major part of the Moon-mass clumps with $M=2\times10^{-8}M_{\odot}$,
$n_p=1$ and $\nu=2$ do not survive inside the central $15$~kpc.

However, in the real Galactic halo the DM clumps have an elongated
orbits in general. These orbits cross the stellar Galactic disk
only {\em once} during the orbital period. Therefore, we must
introduce a factor 2 in (\ref{tdisk}) for these orbits. At the
same time it is a much more important that clumps with elongated
orbits have the longer orbital periods $T_c$ than in the previous
toy model. This elongation of orbits significantly increases the
probability of clump survival.

As an example let us consider the Galactic halo model with an isotropic
velocity distribution. This model is appropriate for the halo formed by
the hierarchial clustering of clumps. In this model according to
\cite{Edd16} the energy distribution function $f(E)$ of DM particles is
related with the density profile of the halo $\rho_{\rm H}(r)$ as
\begin{eqnarray}
 \rho_{\rm H}(r)&=&
 2^{5/2}\pi\int\limits_{U(r)}^{0}\sqrt{E-U(r)}\,f(E)\,dE, \\
 f(E)&=&\frac{1}{2^{3/2}\pi^2}\frac{d}{dE}
 \int\limits_{r(E)}^{\infty}\frac{dr}{\sqrt{E-U(r)}}
 \frac{d\rho_{\rm H}(r)}{dr},
 \label{ferho}
\end{eqnarray}
where $U(r)$ is a gravitational potential energy and function $r=r(E)$
is defined by the equation $U[r(E)]=E$.

\begin{figure}[t]
\includegraphics[width=0.48\textwidth]{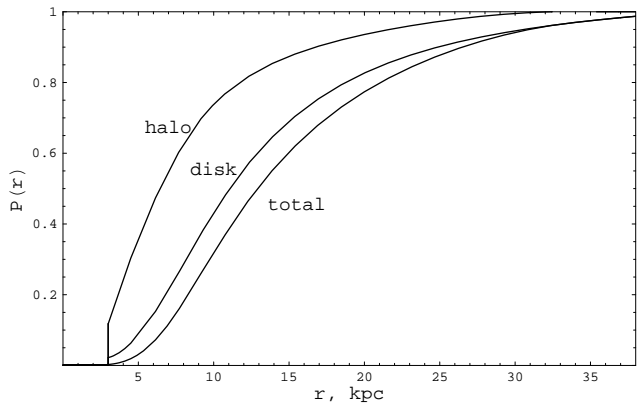}
\caption{The fraction of clumps with mass
$M=2\times10^{-8}M_{\odot}$ and peak-height $\nu=2$ survived a
tidal destruction in the Galactic disc $P_d$, in the Galactic halo
$P_{\rm H}$ and the resulting total fraction $P_{\rm tot}=P_{\rm
H} P_d$ as a function of distance from the Galactic center. The
cutoff at $r<3$~kpc is due to destruction of clumps inside the
bulge. } \label{figdtreal1}
\end{figure}

\begin{figure}[t]
\includegraphics[width=0.48\textwidth]{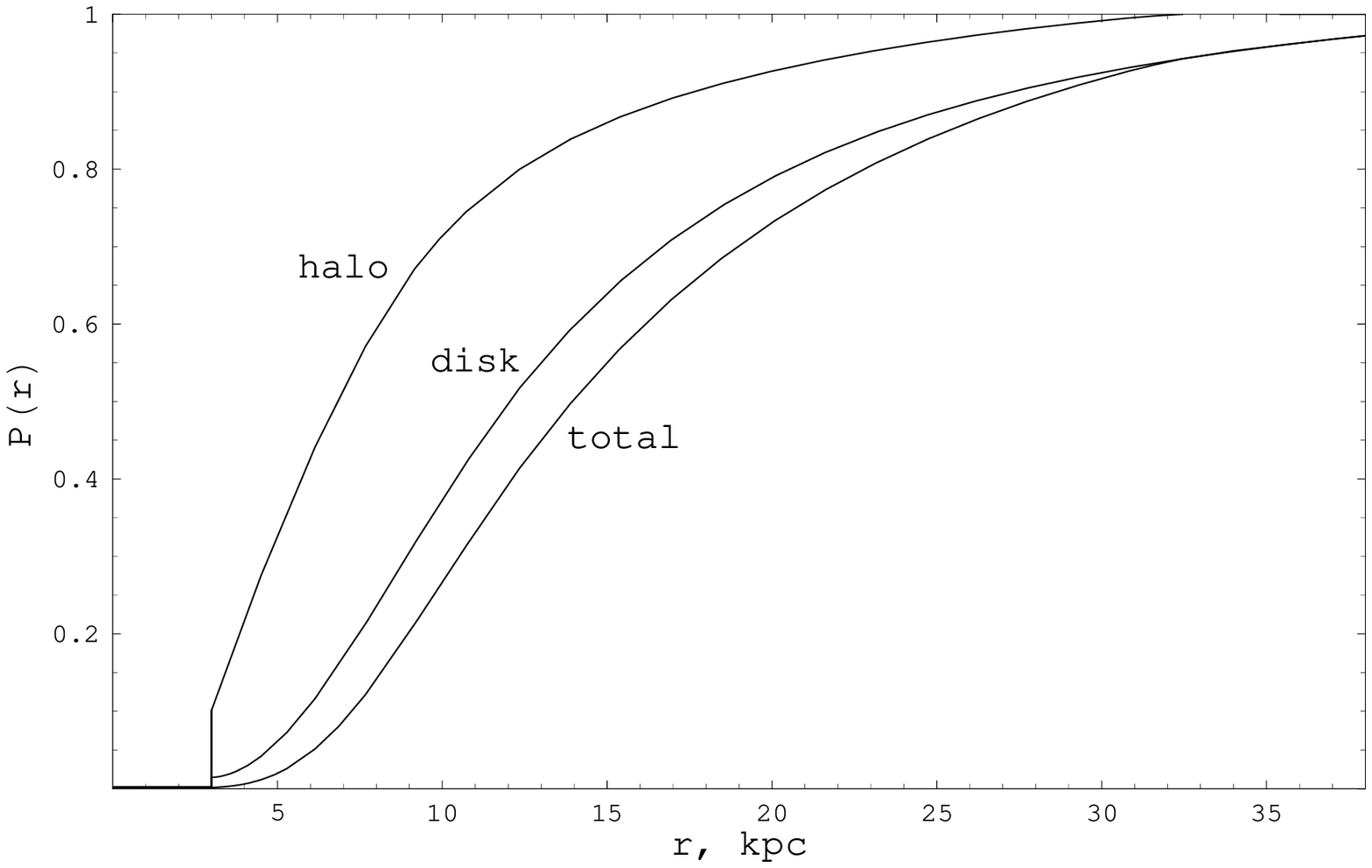}
\caption{The same as the Fig.~\ref{figdtreal1} but for
$M=10^{-6}M_{\odot}$.} \label{figdtreal2}
\end{figure}

\begin{figure}[ht]
\includegraphics[width=0.48\textwidth]{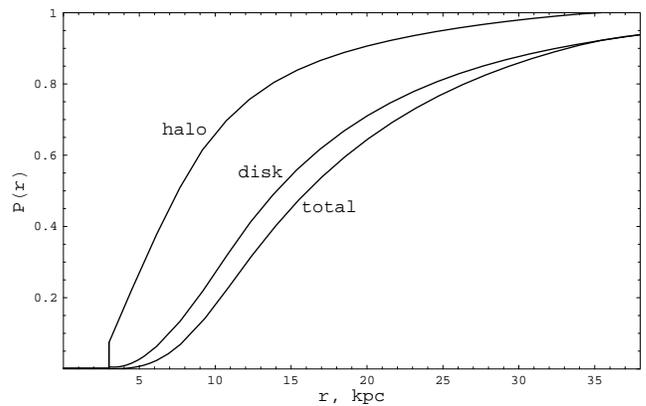}
\caption{The same as the Fig.~\ref{figdtreal1} but for
$M=10^{-3}M_{\odot}$.} \label{figdtreal3}
\end{figure}

We suppose here for simplicity a pure isothermal density profile of the
halo
\begin{equation}
 \rho_{\rm H}(r)=\frac{1}{4\pi}\frac{v_{\rm H}^2}{Gr^2},
 \label{pureiso}
\end{equation}
where $v_{\rm H}=(GM_{\rm H}/R_{\rm H})^{1/2}$ is the halo rotational
velocity and $R_{\rm H}$ is the Galactic halo radius. At $r>R_{\rm H}$
we put $\rho_{\rm H}(r)=0$. In the absence of an analytical model of
the finite isothermal sphere, we construct a simplified model which
approximates the isothermal sphere in the inner region, at $r\ll R_{\rm
H}$. A gravitational potential energy $U(r)$ corresponding to the
density profile (\ref{pureiso}) is
\begin{equation}
 U(r)=mv_{\rm H}^2[\log(r/R_{\rm H})-1],
 \label{pot}
\end{equation}
where $m$ is a mass of DM particle. The radial motion of a particle
with mass $m$ and angular momentum $L$ in the spherical potential obeys
the equation
\begin{equation}
 \dot r^2=\frac{2}{m}\left[E-U(r)\right]-\frac{L^2}{m^2r^2}.
 \label{orbrazm}
\end{equation}
By introducing the dimensionless variables
\begin{equation}
 s=\frac{r}{R_{\rm H}}, \quad
 x=\frac{E}{mv_{\rm H}^2}, \quad
 y=\frac{L^2}{R_{\rm H}^2m^2v_{\rm H}^2},
 \label{dimless}
\end{equation}
the equation for the turning points, $\dot r^2=0$, in the potential
(\ref{pot}) can be written as
\begin{equation}
 \frac{y}{s^2}=2(x-\log s+1).
 \label{minmaxbezr}
\end{equation}
The derivatives of the left and right sides of this equation are equal
at $s=y^{1/2}$. Respectively, the roots $s_{\rm min}(x,y)$ and $s_{\rm
max}(x,y)$ of (\ref{minmaxbezr}) satisfy the condition $s_{\rm
min}(x,y) < y^{1/2} < s_{\rm max}(x,y)$. A condition for the existence
of the solution of (\ref{minmaxbezr}) is $x\ge(\log y-1)/2$. The
equality in this condition corresponds to the circular orbit with
$s_{\rm min}=s_{\rm max}$. From (\ref{orbrazm}) one can determine the
orbital period:
\begin{equation}
 T_c(x,y)=2\frac{R_{\rm H}}{v_{\rm H}}
 \int\limits_{s_{\rm min}}^{s_{\rm max}}
 \frac{ds'}{\sqrt{2(x-\log s'+1)-y/s'^2}}.
 \label{pcint}
\end{equation}
In the following we will solve (\ref{minmaxbezr}) and find the orbital
period $T_c(x,y)$ from (\ref{pcint}) numerically. Denoting
$p=\cos\theta$, where $\theta$ is an angle between the radius-vector
$\vec r$ and the particle velocity $\vec v$, we have
\begin{equation}
 y=2(1-p^2)s^2(x-\log s+1)
 \label{yexp}.
\end{equation}
We find from (\ref{ferho}) the distribution function of particles with
an energy $x<-1$ by using the density profile (\ref{pureiso}) with a
cutoff at $r=R_{\rm H}$:
\begin{equation}
 f(x)=
 \frac{1}{2^{5/2}\pi^3e}
 \frac{v_{\rm H}^{1/2}}{Gm^{3/2}R_{\rm H}^2}\,F(x)
 \label{feerf},
\end{equation}
where
\begin{equation}
 F(x)=\sqrt{2\pi}\,e^{-2x}\,{\rm erf}\left[\sqrt{-2(x+1)}\right]\!+\!
 \frac{e^2}{\sqrt{-(x+1)}}
 \label{jerf}.
\end{equation}
Note that the isotropic distribution function (\ref{feerf}) reproduces
the density profile (\ref{pureiso}) only in the inner halo region, at
$r\ll R_{\rm H}$. The assumed isotropy of particle distribution (i.~e.
an independence of distribution function on the particle angular
momentum $L$) is violated near the boundary of the halo, at $r\simeq
R_{\rm H}$. Nevertheless, the distribution function (\ref{feerf}) is
adequate for our purpose because a tidal destruction of clumps by the
Galactic disk takes place only in the inner halo region, at $r\ll
R_{\rm H}$. At the same time, in the considered model the clumps on the
outer orbits, at $r\simeq R_{\rm H}$, provide only small contribution
to the halo density at $r\ll R_{\rm H}$. Neglecting these outer clumps,
we can define the probability of clump survival in tidal destruction by
the Galactic disk (or the fraction of survived clumps) as a function of
radius $r=sR_{\rm H}$ in the following form:
\begin{equation}
 P_d(r)=\frac{\int\limits_0^1dp\int\limits_{\log
 s-1}^{-1}\!\!dx\,\sqrt{x-\log s+1}\,F(x)\,e^{-t_0/t_d}}
 {\int\limits_{\log s-1}^{-1}\!\!dx\,\sqrt{x-\log s+1}\,F(x)}.
 \label{pbigd}
\end{equation}
Here $t_d$ is from (\ref{tdisk}) but with an additional factor 2 (one
disk crossing per orbital period) and with the replacements
$T_c\Rightarrow T_c(x,y)$ from (\ref{pcint}), $g_m\Rightarrow
g_m(r_{\rm min})$, $v_{z,c}\Rightarrow v(r_{\rm min})$, where $r_{\rm
min}=s_{\rm min}R_{\rm H}$, $s_{\rm min}$ is the minimal root of
(\ref{minmaxbezr}) and $v(r)=\sqrt{2[E-U(r)]/m}$. See in the
Figs.~\ref{figdtreal1}--\ref{figdtreal3} the resulting probabilities of
clump survival in the Galaxy (or fractions of clumps survived the tidal
destruction).

Note that the probability of clump survival $P_d(r)$ in (\ref{pbigd})
is an approximate expression averaged over an angle between the plane
of the Galactic disk and the clump orbit plane. In the real Galactic
halo there must be some anisotropy in clump distribution with respect
to the disk plane. For example, the clumps with orbits in the Galactic
disk plane are destructed more efficiently than ones outside the
Galactic plane.

\section{Destruction of clumps by stars}
\label{stars}

The internal energy increase of a clump during a single star flyby
is
\begin{equation}
\Delta E=\frac{1}{2}\int d^3r\,\rho_{\rm int}(r)(v_z-\tilde v_z)^2,
 \label{tidde}
\end{equation}
where $v_z$ is a velocity increase of constituent DM particle inside a
clump in the direction of axis $z$ and $\tilde v_z$ is a similar one
for a clump center-of-mass. The axis $z$ is directed along the line
connecting a clump center-of-mass with a star at the moment of a star
closest approach. In the impulse approximation, by neglecting the
internal motion of DM particles in a clump during the star encounter
and assuming the straight line orbit of a star (see e.~g.
\cite{gnedin1}) we have
\begin{equation}
 v_z-\tilde v_z\simeq\frac{\partial v_z}{\partial l}\,\Delta l=
 \frac{\partial v_z}{\partial l}\,r\cos\psi,
 \label{tiddvx}
\end{equation}
where $l$ is the distance of a star closest approach to a DM clump and
$\psi$ is a polar angle in the spherical coordinates.

Let us $v_{\rm rel}$ is relative velocity of a star with respect to a
DM clump. In the approximation of a rectilinear motion, an angle $\phi$
between the line connecting a clump center-of-mass and $\vec v_{\rm
rel}$ evolves as
\begin{equation}
\frac{d\phi}{dt}=-\frac{v_{\rm rel}}{l}\cos^2\phi.
 \label{phi}
\end{equation}
Changing a variable $t$ to $\phi$ in the Newton equation of motion, one
gets
\begin{equation}
\frac{dv_z}{d\phi}=-\frac{Gm_*}{v_{\rm rel}l}\,\cos\phi,
\end{equation}
where $m_*$ is a typical star mass. After integration of this equation
we obtain
\begin{equation}
v_z=\frac{2Gm_*}{v_{\rm rel}l}.
\end{equation}
Now by integrating (\ref{tidde}) over a clump volume with a density
profile $\rho_{\rm int}(r)$ from (\ref{rho}), we find  in the case of
$l>R$:
\begin{equation}
 \Delta E= \frac{2(3-\beta)}{3(5-\beta)}
 \frac{G^2MR^2m_*^2}{v^2_{\rm rel}l^4}.
\end{equation}
The opposite case $l<R$ was considered e.~g. in \cite{bde03}. It is
easily verified that the maximum internal energy increase occurs for a
star flyby with $l\simeq R$.

\begin{figure}[t]
\includegraphics[width=0.48\textwidth]{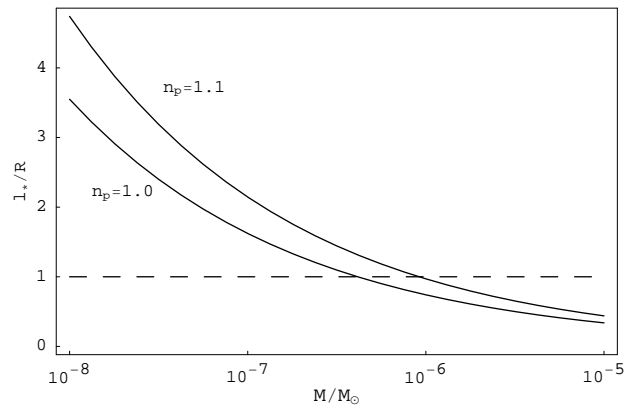}
\caption{The fraction $l_*/R$ according to (\ref{eqlzv}) as a function
of clump mass $M$ at the distance from the Galactic center $r=2$~kpc,
$\nu=2$ and $n_p=1.0$ and $n_p=1.1$ respectively.}
 \label{lzvr}
\end{figure}
At this step we must distinguish two cases: (i) clump destruction
during a single star flyby and (ii) clump destruction after numerous
star collisions. In the first case a threshold for clump destruction is
achieved at $\Delta E=|E|$, where a total energy of clump $E$ is given
by (\ref{Etot}). From the equality $\Delta E=|E|$ one finds the maximal
impact parameter $l_*$ for a single flyby destruction
\begin{equation}
 \left(\frac{l_*}{R}\right)^4=
 \frac{4(5-2\beta)}{3(5-\beta)} \frac{Gm_*^2}{MRv_{\rm rel}^2}
 \sim\left(\frac{V}{v_{\rm rel}}\right)^2\left(\frac{m_*}{M}\right)^2,
 \label{eqlzv}
\end{equation}
where $V\simeq(GM/R)^{1/2}$ is a velocity dispersion of DM particles in
the clump. The fraction $l_*/R$ as a function of clump mass $M$ is
shown in the Fig.~\ref{lzvr}. Note that condition $l_*/R>1$ is
satisfied for clumps of the smallest mass. A total rate of clump
destruction by stars in the case of $l_*/R>1$ is given by
\begin{equation}
 t_*^{-1}=\frac{\dot E}{|E|}=
 \pi l_*^2n_*v_{\rm rel}+\frac{\dot E(l>l_*)}{|E|}.
 \label{nud}
\end{equation}
where $n_*$ is a number density of stars and
\begin{equation}
 \dot E(l>l_*)=2\pi
 \int\limits_{l_*}^{\infty}\!\Delta E(l)\,n_*v_{\rm rel}\,l\,dl.
 \label{dee1}
\end{equation}
After integration in (\ref{dee1}) with $l_*$ from (\ref{eqlzv}), we
find that the second term in (\ref{nud}) is equal to the first one.
Thus, the resulting time of clump destruction in the case of $l_*/R>1$
is
\begin{equation}
 t_{*}=\frac{1}{2\pi l_*^2v_{\rm rel}n_*}=
 \frac{1}{4\pi n_*m_*}
 \left[\frac{3(5-\beta)}{(5-2\beta)}\frac{M}{GR^3}\right]^{1/2}\!.
  \label{td1}
\end{equation}
We see from (\ref{td1}) that the time of clump destruction by stars
does not depend on $v_{\rm rel}$ in the case of $l_*/R>1$.

Similarly, the time of clump destruction by stars in the case of
$l_*<R$ is
\begin{equation}
 t_{*}=\frac{3(5-\beta)}{8\pi(5-2\beta)}
 \frac{v_{\rm rel}M}{GRm_*^2n_*}.
 \label{td2}
\end{equation}

\subsection{Destruction of clumps in the bulge}

The bulge is central spheroidal subsystem of the Galaxy. Following
to \cite{LauZylMez} we approximate the radial number density
distribution of stars in the bulge in the radial range $r=1-3$~kpc
as
\begin{equation}
 n_{b,*}(r)=(\rho_b/m_*)\exp\left[ -(r/r_b)^{1.6}\right],
 \label{rhoe}
\end{equation}
where $\rho_b=8M_{\odot}/$pc$^3$ and $r_b=1$~kpc. By using
(\ref{rhoe}) together with (\ref{td1}) or (\ref{td2}) it can be
shown that inside the bulge, at $r\leq3$~kpc, all small-scale
clumps with $M\geq10^{-8}M_{\odot}$ are tidally destructed during
the Hubble time, i.~e. $t_*\ll t_0$. Therefore, there is an empty
cavity in clump distribution in the Galactic center with a radius
$r\simeq3$~kpc as it can be shown in the
Figs.~\ref{figdtreal1}--\ref{figdtreal3}.

What is a fate of the core of a tidally destructed clump? Let us
consider the scaling of destruction time $t_*$ in dependence on a
varying clump radius $r$ and mass $M(r)\propto r^{3-\beta}$. According
to (\ref{td1}) or (\ref{td2}), a time of clump destruction is scaled
respectively as $t_*\propto r^{-\beta/2}$ or $t_*\propto r^{2-\beta}$.
For surviving of the core, it is necessary that $t_*\to0$ at $r\to0$.
This is possible only if the internal density profile $\rho(r)\propto
r^{-\beta}$ is rather steep, $\beta>2$. Meanwhile, both the theoretical
models and numerical simulations predict $\beta<2$, and therefore, the
core does not survive during the tidal destruction of DM clump.

\subsection{Destruction of clumps in the halo}

The radial number density distribution of stars in the Galactic halo
(outside the Galactic disk) at radii $r>3$~kpc can be approximated as
\begin{equation}
  n_{h,*}(r)=(\rho_\odot/m_*) (r_{\odot}/r)^{3},
 \label{rhosh}
\end{equation}
where $\rho_{\odot}=10^{-4}~M_{\odot}/$pc$^3$ and $r_{\odot}=8.5$~kpc.
The stellar density profile in the Galactic halo is rather poorly
known, and so (\ref{rhosh}) must be considered only as an upper limit
\cite{MDSQ}. We will describe the distribution of clumps in the halo
the same way as in the Section~\ref{disksec}. To take into account a
varying number density of stars $n_{h,*}(r)$, we made an averaging of
the rate of clump destruction by stars $t^{-1}_*$ along the orbital
trajectory during the orbital period:
\begin{equation}
 \langle t_*^{-1}(x,y)\rangle\!=
 \!\frac{2R_{\rm H}}{v_{\rm H}T_c(x,y)}
 \!\int\limits_{s_{\rm min}}^{s_{\rm max}}\!\!
 \frac{ds'\,t_*^{-1}}{\sqrt{2(x-\log s'+1)-y/s'^2}}.
 \label{pcintav}
\end{equation}
In this expression the dimensionless variables $x$ and $y$ are
from (\ref{dimless}), a clump orbital period $T_c(x,y)$ is from
(\ref{pcint}) and destruction time of clump $t_*$ is from
(\ref{td1}) or (\ref{td2}) with the replacements $n_s\Rightarrow
n_{h,*}(r)$ and $v_{\rm rel}\Rightarrow \sqrt{2[E-U(r)]/m}$.  The
averaging procedure (\ref{pcintav}) is in fact the integration of
energy gain rate $\int\dot Edt$ along the clump orbit.

The resulting probability of clump survival in tidal destruction by the
Galactic halo stars $P_{\rm H}(r)$ is defined by a similar expression
as (\ref{pbigd}) but with a replacement $e^{-t_0/t_d}\Rightarrow
e^{-t_0\langle t_*^{-1}(x,y)\rangle}$, where $\langle
t_*^{-1}(x,y)\rangle$ is from (\ref{pcintav}).

The results of numerical calculations of the tidal destruction of
DM clumps by different Galactic components are summarized in the
Figs.~\ref{figdtreal1}--\ref{figdtreal3}. These calculations were
performed for DM clumps originated from fluctuations with the
peak-height $\nu=2$.

Correspondingly 32\%, 27\% and 18\% of clumps survive the destruction
by the Galactic disk tidal shocking at the Sun position,
$r_{\odot}=8.5$~kpc, for clump masses $M=2\times10^{-8}M_{\odot}$,
$M=10^{-6}M_{\odot}$ and $M=10^{-3}M_{\odot}$. The  Galactic disk
destroys clumps even outside its boundary, at $r>15$~kpc, because some
of DM clumps with the extended orbits intersect the Galactic disk in
the inner part of the halo. The destruction of clumps by the Galactic
disk becomes inefficient at $r>40$~kpc. The respective fractions of
clumps of the same masses surviving the tidal destruction by stars in
the Galactic halo (outside the Galactic disk) are 66\%, 63\% and 57\%.
The final fractions of clumps of the same masses survived the tidal
destruction both by the Galactic disk and stars in the Galactic halo
$P(r_{\odot})= P_{\rm H}(r_{\odot}) P_d(r_{\odot})$ are 21\%, 17\% and
10\% respectively.

\begin{figure}
\includegraphics[width=0.48\textwidth]{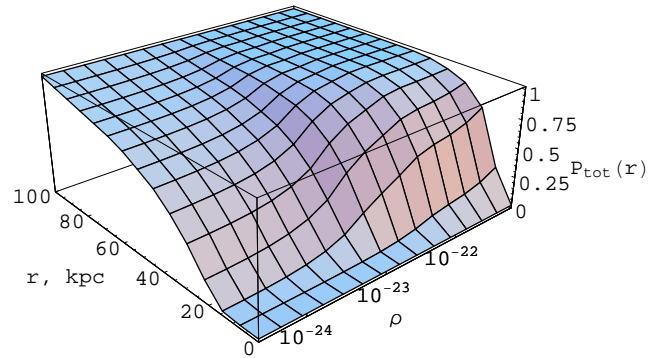}
\caption{The survived fraction of small-scale clumps $P_{\rm
tot}(r)$ in the Galactic halo inside the radial distance
$r\leq100$~kpc. A mean internal density of clump $\rho$ is in
g~cm$^{-3}$. } \label{prf100prf15}
\end{figure}

In the Fig.~\ref{prf100prf15} the fraction of survived small-scale
clumps in the Galactic halo is shown in dependance on a mean internal
density of small-scale clumps.

\section{Conclusions}
\label{discussion}

We calculated the number density distribution of small-scale DM
clumps in the Galactic halo in dependance on a clump mass $M$,
radius $R$ (expressed through the fluctuation peak-height $\nu$)
and radial distance $r$ to the Galactic center. These calculations
were performed by taking into account the tidal destruction of
clumps in the early hierarchical clustering and later in the
Galaxy.

Calculations of the distribution function of small-scale clumps are
carried out, following to \cite{bde03}, in the framework of the
standard cosmological model and the hierarchical model of structure
formation. The primeval power spectrum of density perturbations $P(k)
\propto k^{n_p}$ is taken from the inflation models with $n_p \simeq 1$
(the Harrison-Zeldovich spectrum). In this model the small-scale clumps
are formed earlier than the bigger ones. The minimal mass $M_{\rm min}$
of clumps is determined by the free streaming of DM particles from a
growing fluctuation. The value of $M_{\rm min}$ is a model dependent
quantity. For neutralino as DM particle, the minimal mass $M_{\rm min}$
is given by (\ref{mminnum}), and it is the Moon-scale mass.

In the process of hierarchical clustering the small clumps are
captured by the bigger ones, and so on. Thus the hierarchical
structure is formed, when all clumps consist in part the smaller
ones and the free DM particles. Some part of DM clumps are tidally
disrupted in the gravitational field of the bigger host clumps. In
this scenario we calculated the differential distribution of the
survived clumps given by (\ref{psiitog}) as a function of two
independent parameters: e.~g. a clump mass $M$ and fluctuation
peak-height $\nu$ (or a clump mass $M$ and radius $R$). The
corresponding integral mass function is given by (\ref{xitot}),
where the small factor $\xi_{\rm int} \simeq 0.017(n+3)$ gives the
mass fraction of clumps survived the tidal destruction in the
hierarchical structuring.

The predicted differential number density of small clumps
$\xi(M)\,(\rho_{\rm H}/M)\,dM/M$ is very close to our previous
calculations \cite{bde03}, and both are in a good agreement with the
recent results of numerical simulations \cite{DieMooSta05}.

Our calculations are valid only for small-scale clumps with masses
$M\le 10^3M_{\odot}$. The physics of larger mass clumps is rather
different. For large-scale clumps the dynamical friction, tidal
stripping and accretion of new clumps into the halo proceed in a
different way. Nevertheless, the calculated mass function is in a good
agreement with a mass function of the large clumps (obtained in the
numerical simulations) in the intermediate mass range (see in the
Fig.~\ref{figsh}).

The mutual tidal destruction of small-scale DM clumps is effective only
at the early stage of hierarchical clustering. At later stages the DM
clumps are additionally destructed by stars and by the collective
gravitational field of the Galactic disc. In the Galaxy at radial
distance $r \leq 3$~kpc all small-scale clumps are destructed by stars
in the central bulge. At radial distances in the range $r=3-40$~kpc the
DM clumps are destructed by stars from the halo and by the tidal
shocking in the Galactic disk. The latter provides the major
contribution to the tidal destruction of clumps outside the bulge. Only
21\%, 17\% and 10\% of clumps survive the tidal destruction near the
Sun position for clump masses $M=2\times10^{-8}M_{\odot}$,
$M=10^{-6}M_{\odot}$ and $M=10^{-3}M_{\odot}$ respectively. Our results
on the tidal destruction of clumps differ from both
\cite{ZTSH,ZTSH0508} and \cite{MDSQ}, with the intermediate
conclusions. At radial distances $r>40$~kpc the destruction of clumps
by the Galactic disk becomes inefficient, and the number density of
clumps is determined only by the early epoch of hierarchical
clustering.

The tidal destruction of clumps by the Galactic disk and stars
affects the annihilating signal mainly in the central region of
the Galaxy where destructions are most effective. Therefore, a
growing fraction of survived clumps $P(r)$ smooths the anisotropy
of the awaited annihilation signal at the Sun position. A local
annihilation rate is proportional to the clumps number density
and, respectively, to $P(r)$. For example, at the position of the
Sun the 17\% of clumps survive, and so the local annihilation rate
more then 5 times less in comparison with the $P=1$ case.

\begin{acknowledgments}
We acknowledge the anonymous referee for valuable comments. We thank
ILIAS-TARI for access to the LNGS research infrastructure and for the
financial support through EU contract RII3-CT-2004-506222. This work
has been supported in part by the Russian Foundation for Basic Research
grants 02-02-16762, 03-02-16436, and 04-02-16757 and by the Ministry of
Science of the Russian Federation, grants 1782.2003.2 and 2063.2003.2.
\end{acknowledgments}

\end{document}